\documentclass[12pt]{article} 
\usepackage{ucs} 
\usepackage[utf8x]{inputenc} 
\usepackage{amsmath,amssymb} 
\usepackage[nointegrals]{wasysym} 
\usepackage{graphicx} 
\usepackage{color} 
\usepackage{hyperref} 
\hypersetup{bookmarksopen=true, bookmarksnumbered=true, pdfstartview={FitH}, pdfborder={0 0 0}, colorlinks=true, linkcolor=red, linktocpage=true, citecolor=blue, urlcolor=cyan, unicode=true, pdftitle={CS A SYM}, pdfauthor={DJ}, pdfsubject={N=2 SYM Action}, pdfkeywords={Π, }{♫, }{CS, }{SYM}}
\usepackage{geometry} 
\usepackage{fancyhdr} 
\usepackage{parskip} 
\usepackage[nottoc,notlof,notlot]{tocbibind} 
\usepackage[titles]{tocloft} 

\definecolor{title}{rgb}{0.3,0.3,0.9}
\definecolor{abst}{rgb}{0.366,0.366,0.266}
\definecolor{sect}{rgb}{1.0,0.0,0.0}
\definecolor{ssect}{rgb}{0.5,0.5,0.0}
\definecolor{sssect}{rgb}{0.3,0.3,0.3}
\definecolor{appsect}{rgb}{0.0,1.0,0.0}
\definecolor{ref}{rgb}{0.0,0.0,1.0}

\newcommand\sect[1] {{\color{sect}\section{#1}}}
\newcommand\subsect[1] {{\color{ssect}\subsection{#1}}}

\newcommand\references[1] {\color{ref} \begin{thebibliography}{[99]} \color{black} #1 \end{thebibliography}}
\numberwithin{equation}{section} 

\newcommand\bc {\begin{center}}
\newcommand\ec {\end{center}}
\newcommand\bnum {\begin{enumerate}}
\newcommand\enum {\end{enumerate}}
\newcommand\be {\begin{equation}}
\newcommand\ee {\end{equation}}
\newcommand\ben {\begin{equation*}}
\newcommand\een {\end{equation*}}
\newcommand\bfig {\begin{figure}}
\newcommand\efig {\end{figure}}
\newcommand\bpm {\begin{pmatrix}}
\newcommand\epm {\end{pmatrix}}
\renewcommand\({\left(}
\renewcommand\){\right)}
\renewcommand{\exp}{e^}
\renewcommand{\i}{\dot{\iota}}
\newcommand\D {{\cal D}}
\renewcommand\P {{\cal P}}
\renewcommand\S {{\cal S}}
\renewcommand\' {^{\prime}}

\DeclareUnicodeCharacter{"0393}{\Gamma}
\DeclareUnicodeCharacter{"0394}{\Delta}
\DeclareUnicodeCharacter{"0398}{\Theta}
\DeclareUnicodeCharacter{"039B}{\Lambda}
\DeclareUnicodeCharacter{"039E}{\Xi}
\DeclareUnicodeCharacter{"03A0}{\Pi}
\DeclareUnicodeCharacter{"03A3}{\Sigma}
\DeclareUnicodeCharacter{"03A5}{\Upsilon}
\DeclareUnicodeCharacter{"03A6}{\Phi}
\DeclareUnicodeCharacter{"03A8}{\Psi}
\DeclareUnicodeCharacter{"03A9}{\Omega}
\DeclareUnicodeCharacter{"03B1}{\alpha}
\DeclareUnicodeCharacter{"03B2}{\beta}
\DeclareUnicodeCharacter{"03B3}{\gamma}
\DeclareUnicodeCharacter{"03B4}{\delta}
\DeclareUnicodeCharacter{"03B5}{\epsilon}
\DeclareUnicodeCharacter{"03B6}{\zeta}
\DeclareUnicodeCharacter{"03B7}{\eta}
\DeclareUnicodeCharacter{"03B8}{\theta}
\DeclareUnicodeCharacter{"03D1}{\vartheta}
\DeclareUnicodeCharacter{"03B9}{\iota}
\DeclareUnicodeCharacter{"03BA}{\kappa}
\DeclareUnicodeCharacter{"03BB}{\lambda}
\DeclareUnicodeCharacter{"03BC}{\mu}
\DeclareUnicodeCharacter{"03BD}{\nu}
\DeclareUnicodeCharacter{"03BE}{\xi}
\DeclareUnicodeCharacter{"03C0}{\pi}
\DeclareUnicodeCharacter{"03C1}{\rho}
\DeclareUnicodeCharacter{"03C3}{\sigma} 
\DeclareUnicodeCharacter{"03C4}{\tau}
\DeclareUnicodeCharacter{"03C5}{\upsilon}
\DeclareUnicodeCharacter{"03C6}{\phi}
\DeclareUnicodeCharacter{"03D5}{\varphi}
\DeclareUnicodeCharacter{"03C7}{\chi}
\DeclareUnicodeCharacter{"03C8}{\psi}
\DeclareUnicodeCharacter{"03C9}{\omega}
\DeclareUnicodeCharacter{"21D0}{\Leftarrow}
\DeclareUnicodeCharacter{"0212B}{\AA}
\DeclareUnicodeCharacter{"266A}{\eighthnote}
\DeclareUnicodeCharacter{"266B}{\twonotes}

\newcommand\arXivid[1] {\href{http://arxiv.org/abs/#1}{\tt arXiv:#1}} 
\newcommand\cmp[3] {{\it Commun.\ Math.\ Phys.\ }\href{http://inspirehep.net/search?ln=en&ln=en&p=find+j+"Commun.Math.Phys.,#1,#3"&of=hb&action_search=Search&sf=&so=d&rm=&rg=25&sc=0}{{\bf #1} (#2) #3}} 
\newcommand\cqg[3] {{\it Class.\ Quant.\ Grav.\ }\href{http://inspirehep.net/search?ln=en&ln=en&p=find+j+"Class.Quant.Grav.,#1,#3"&of=hb&action_search=Search&sf=&so=d&rm=&rg=25&sc=0}{{\bf #1} (#2) #3}}
\newcommand\jhep[3]{{\it JHEP\ }\href{http://inspirehep.net/search?ln=en&ln=en&p=find+j+"JHEP,#1,#3"&of=hb&action_search=Search&sf=&so=d&rm=&rg=25&sc=0}{{\bf #1} (#2) #3}}
\newcommand\npb[3] {{\it Nucl.\ Phys.\ }{\bf B #1} (#2) #3}
\newcommand\pr[4] {{\it Phys.\ Rev.\ }{\bf #1 #2} (#3) #4} 

\begin{document}
\pagenumbering{alph}
\title{\color{title}\Huge N=2 Super-Yang-Mills Theory from a Chern-Simons Action}
\author{Dharmesh Jain\footnote{\href{mailto:djain@insti.physics.sunysb.edu}{djain@insti.physics.sunysb.edu}}\, , Warren Siegel\footnote{\href{mailto:siegel@insti.physics.sunysb.edu}{siegel@insti.physics.sunysb.edu}\vskip 0pt \hskip 8pt \href{http://insti.physics.sunysb.edu/\~siegel/plan.html}{http://insti.physics.sunysb.edu/$\sim$siegel/plan.html}}\bigskip\\ \emph{C. N. Yang Institute for Theoretical Physics}\\ \emph{State University of New York, Stony Brook, NY 11794-3840}}
\date{} 
\maketitle
\thispagestyle{fancy}
\rhead{YITP-SB-12-05} 
\lhead{\today}
\begin{abstract}
\normalsize We present a Chern-Simons action for N=2 Super-Yang-Mills theory (SYM) in `full' N=2 superspace (hyperspace) augmented by coordinates of the internal SU(2) group and show that this action can be reduced to the usual SYM action in the Harmonic ($♫$) hyperspace. We also discover that the `choice' of Harmonic hyperspace is not unique and under suitable conditions, further reduction to the well-known Projective ($\check{Π}$) hyperspace is possible.
\end{abstract}

\tableofcontents

\newpage
\pagenumbering{arabic}
\cfoot{\thepage}\rhead{}\lhead{}
\sect{Introduction}
The harmonic hyperspace ($♫$) formalism was developed by GIKOS in \cite{GIKOS2-1}, which allowed writing down the actions for various N=2 supermultiplets. Specifically, the action for non-Abelian super-Yang-Mills (SYM) multiplet was written as an infinite series expansion in terms of the prepotential. This action also turns out to be non-local in the internal R-coordinates. Even though the origin of the Abelian action could be understood via the action written in chiral hyperspace, the non-Abelian action did not have such a direct origin. Its origin was explained by Zupnik in \cite{Zup} where the `series' action was summed to a logarithm of a pseudo-differential operator. 

In this paper, we present a different origin for the non-Abelian SYM action. We show that a Chern-Simons (CS) action for SYM can be written in `full' hyperspace ($d^4x\,d^8θ$) supplemented by the internal SU(2) space ($d^3y$). However, as the CS action doesn't know about the geometry of the space, we can choose a `different' internal space as long as integration over this space can be consistently defined. Thus, choosing a space with a boundary (amounts to a suitable Wick-rotation of SU(2)) is desirable as this (local) CS action can then be `reduced' to the (non-local) SYM action of $♫$ on this boundary. This also means that the sphere is not the only possibility for the harmonic internal space and other spaces can be chosen as we'll see in section \ref{CSHH}, which facilitate further reduction to projective hyperspace ($\check{Π}$).

We have shown in \cite{DPHfH} that the $♫$ can be reduced to $\check{Π}$ (developed by Lindström and Roček\cite{ULMR}) after Wick rotating the internal 2-sphere and restricting the dynamics of hyperfields to one of the resultant boundaries. Here, we modify the arguments slightly to make the reduction more concrete and we will see that we get to the same $\check{Π}$.

\sect{Review of SYM in Harmonic Hyperspace}
We use the notations of \cite{WS10} to denote the coordinates and derivatives. The eight Fermionic coordinates (apart from the spacetime coordinates $x$), making up the `full' superspace, are labelled as $\{θ^α,\bar{θ}^{\dot{α}}\}$ \& $\{ϑ^α,\bar{ϑ}^{\dot{α}}\}$. The internal symmetry group SU(2) is parameterized by three  Bosonic coordinates denoted by `flat' indices $\{+,-,0\}$. The constraints $d_{ϑ}Φ=\bar{d}_{ϑ}Φ=0$ define a harmonic (analytic) hyperfield `$Φ$'.

The following constraints then define the SYM in $♫$:
{\allowdisplaybreaks
\begin{align}
\{\D_{ϑ},\D_{ϑ}\}=\{\D_{ϑ},\bar{\D}_{ϑ}\}=\{\bar{\D}_{ϑ},\bar{\D}_{ϑ}\}&=0\label{harm}\\
[\D_{+},\D_{ϑ}\(\bar{\D}_{ϑ}\)]&=\D_{θ}\(\bar{\D}_{θ}\)\label{just}\\
[\D_{+},\D_{θ}\(\bar{\D}_{θ}\)]&=0\label{ythe}\\
[\D_{-},\D_{ϑ}\(\bar{\D}_{ϑ}\)]&=0\label{prep}\\
[\D_{-},\D_{θ}\(\bar{\D}_{θ}\)]&=\D_{ϑ}\(\bar{\D}_{ϑ}\)\label{ybth}\\
[\D_{0},\D_{ϑ}\(\bar{\D}_{ϑ}\)]&=-\D_{ϑ}\(\bar{\D}_{ϑ}\)\label{0bth}\\
[\D_{0},\D_{θ}\(\bar{\D}_{θ}\)]&=\D_{θ}\(\bar{\D}_{θ}\)\label{0the}\\
[\D_{-},\D_{+}]&=2\D_{0}\label{ssol}\\
[\D_{0},\D_{\pm}]&=\pm\D_{\pm}\label{0yyb}
\end{align}}
where $\D$'s are gauge covariant derivatives: $\D=d+A$. In $♫$, the coordinate denoted by `$0$' corresponds to U(1) in the coset SU(2)/U(1) due to which the corresponding derivative is not covariantized, i.e. $A_0=0$. The above constraints are then solved in the following way to get the SYM action:
\bnum
\item Choose the gauge ($λ$-frame): $A_{ϑ}=\bar{A}_{ϑ}=0$.
\item $A_{-}$ becomes a harmonic hyperfield due to equation \ref{prep}. It is also the `prepotential'.
\item Eq. \ref{just} just gives $A_θ=-d_{ϑ}A_{+}$.
\item $A_{+}$ is solved as a series in terms of $A_{-}$ from equation \ref{ssol}:
\be
A_{+}=\sum_{n=1}^{\infty}\(\prod_{i=1}^{n}\int d^2y_i\)\frac{A_{1-}\,...\,A_{n-}}{(y-y_1)y_{12}...(y_n-y)}.
\ee
where $d^2y$ is the volume element of S$^2$, $A_{i-}\equiv A_{-}(x,θ,ϑ,y_i)$ and $y_{12}=(y_1-y_2)$ with a relevant $ε-$prescription defined later.
\item The Abelian action is written as $A_{-}A_{+}$ (derived from the chiral version) which is generalized in the non-Abelian case to a series with an extra factor of $\frac{1}{n}$:
\be
\S_{♫}=-\frac{tr}{2g^2}\int dx\,d^8\theta\sum_{n=2}^{\infty}\frac{(-\i)^n}{n}\(\prod_{i=1}^{n}\int d^2y_i\)\frac{A_{1-}A_{2-}\,...\,A_{n-}}{y_{12}\,y_{23}\,...\,y_{n1}}.
\label{SHH}
\ee
\enum

\sect{Chern-Simons Action for N=2 SYM}
We now work with `curved' SU(2) derivatives, $∂_m (m=1\,,2\,,3)$ instead of the `flat' ones, $d_a (a=+\,,-\,,0)$ used above:
\begin{gather}
d_a={e_a}^m∂_m\nonumber\\
\left[d_a,d_b\right]={f_{ab}}^c d_c \rightarrow [∂_m,∂_n]=0\label{tor0}
\end{gather}
where ${f_{ab}}^c$'s are the SU(2) structure constants (can be read from equations \ref{ssol} \& \ref{0yyb}) and we require that ${e_a}^m$ is a dreibein satisfying
\be
{e_-}^{1}={e_-}^{3}=0.
\label{ec}
\ee

Introducing gauge covariant derivatives $∂_m \rightarrow ∇_m=∂_m+A_m$ in equation \ref{tor0}, we get:
\be
[∇_m,∇_n]=F_{mn}=0\label{eom}
\ee
Let us now check how the spinorial covariant derivatives act on the `curved' SU(2) connections (conjugate derivatives give similar results):
\begin{gather}
[\D_a,\D_{ϑ}]={f_{aϑ}}^η \D_η\rightarrow [∇_m,\D_{ϑ}]={e_m}^a {f_{aϑ}}^η \D_η\\
\Rightarrow d_{ϑ}A_{2}=0 \label{projA}
\end{gather}
where the non-zero constants are: ${f_{0θ}}^θ=-{f_{0ϑ}}^ϑ={f_{+ϑ}}^θ={f_{-θ}}^ϑ=1$ (read from eqs. \ref{prep}$-$\ref{0the}), which imply $A_2$ is a harmonic hyperfield. This result is valid in general due to the condition in equation \ref{ec}.

Finally, the constraints in equation \ref{eom} can be derived as equations of motion from a CS action:
\be
\S_3=\frac{tr}{2g^2}\int dx\,d^8θ\,d^3y\,ε^{mnp}\left[\frac{1}{2}A_m∂_n A_p+\frac{1}{3}A_m A_n A_p\right].\label{CSA}
\ee
This action is reminiscent of the N=3 SYM action in $♫$\cite{GIKOS3-1}. An important difference is that while in the case of N=3 all the $A$'s are `harmonic', only one of them is in N=2 SYM and the above action has `full' hyperspace measure with 8 Fermionic coordinates whereas the N=3 SYM action has only harmonic superspace measure also with 8 $θ$'s instead of all the twelve. Also, the $y-$integration in \ref{CSA} is over three real coordinates corresponding to SU(2)=S$^3$ (or its Wick-rotated versions) whereas for N=3 SYM, the integration is over three complex coordinates corresponding to the coset SU(3)/U(1)$^2$.

\vspace{20pt}
\sect{Reduction from Full Hyperspace to Harmonic\label{CSHH}}
As mentioned in the introduction, we are not restricted to use the compact SU(2) manifold as the internal 3-manifold for the CS action since the geometry does not affect it. Hence, we can choose the internal 3-manifold for the CS action to have a boundary at $y^3=0$, which basically amounts to a `Wick-rotation' of SU(2) to SU(1,1). We do not put any boundary conditions on $A$ at this boundary due to which the variation of action \ref{CSA} reads:
\be
δ\S_3=\frac{tr}{4g^2}\int dx\,d^8θ\,d^3y\,ε^{mnp}\left[2δA_mF_{np}-∂_m\left(A_nδA_p\right)\right].
\ee
The first term gives the usual equations of motion and the second (boundary) term breaks gauge invariance in general\footnote{The action can be made gauge invariant by imposing suitable boundary conditions on $A$ or by adding additional boundary degrees of freedom as shown in \cite{WEMSS, CCDS}. The gauge invariance can also be retained if we allow the gauge parameter to vanish at the boundary, but we do not do that.}. Ignoring this subtlety, we can rewrite the action \ref{CSA} as:
\be
\S_3=\frac{tr}{4g^2}\int dx\,d^8θ\,d^3y\,ε^{ij}\left[2F_{ij}A_3-A_i∂_3A_j-∂_i\left(A_jA_3\right)\right],\label{CSAg}
\ee
where $i=1,2$. The total derivative term vanishes as there are no boundaries along $y^i$. Here, $A_3$ acts as a Lagrange multiplier and being an unconstrained hyperfield imposes the constraint $F_{12}=0$, whose solution can be substituted back to get a simplified action\footnote{Usually, the connections $A_i$ are chosen to be flat at this point and written as $A_i=\(∂_iU\)U^{-1}$, which gives the well-know Wess-Zumino action.}. In other words, we substitute the solution of the equation of motion of $A_3$ so that the action has only harmonic hyperfields:
\be
F_{12}=∂_1 A_2-∂_2 A_1+[A_1,A_2]=0\quad \& \quad A_1=\sum_n A_1^{(n)}\label{F12}
\ee
\begin{align*}
\Rightarrow&\,A_1^{(1)}(y)=∂_1\int d^2 y\' \frac{A_2\'(y\')}{y^1-{y^1}\'+\frac{ε}{{y^2}\'-y^2}}=-\int d^2 y\' \frac{A_2\'(y\')}{\left(y^1-{y^1}\'\right)^2},\\
{}&\,A_1^{(2)}(y)=-\int d^2 y\' d^2 y^{\prime\prime} \frac{A_2\'(y\') A_2^{\prime\prime}(y^{\prime\prime})}{\left(y^1-{y^1}^{\prime\prime}\right)\left({y^1}^{\prime\prime}-{y^1}\'\right)\left({y^1}\'-y^1\right)} , \quad\mathrm{and\, so\, on...}
\end{align*}
\be
\Rightarrow A_1=\sum_{n=1}^{∞}(-1)^{n+1}\int d^2y\'...\,d^2y^{(n)'}\frac{ A_2\'\,...\,A_2^{(n)'}}{(y^1-{y^1}\')\,...\,\left({y^1}^{(n)'}-y^1\right)} \label{A1}
\ee
where $d^2y\equiv dy^1dy^2$ and the $ε-$term is present in all denominator factors. The following identity is used to prove that the solution in \ref{A1} indeed makes the curvature vanish (equation \ref{F12}):
\be
∂_2\left(\frac{1}{{y^1}\'-y^1+\frac{ε}{{y^2}\'-y^2}}\right)\sim δ^2(y\'-y).
\label{delta}
\ee

Plugging this solution back in action \ref{CSAg}, we get:
\begin{align}
\S_3&=-\frac{tr}{4g^2}\int dx\,d^8θ\,d^2 y\,dy^3\left(A_1∂_3A_2-A_2∂_3A_1\right)\label{still3}\\
&=-\frac{tr}{2g^2}\int dx\,d^8θ\int_0^∞dy^3\sum_{n=2}^{∞}\frac{(-1)^n∂_3}{n}\left(\int d^2y\,d^2y\'...\,d^2y^{(n-1)'}\right.\times\nonumber\\
&\qquad\qquad\qquad\qquad\qquad\qquad\times\left.\frac{A_2\, A_2\'\,...\,A_2^{(n-1)'}}{(y^1-{y^1}\')\,...\,\left({y^1}^{(n-1)'}-y^1\right)}\right)\label{almost2}
\end{align}
The equation \ref{almost2} can be written with the factor $\frac{1}{n}$ because all $A_2$'s depend on same $y^3$ (i.e. no primes). Assuming $A_2$ is well-behaved at $y^3=∞$, we can integrate over $y^3$ and write a `2D' action on the boundary at $y^3=0$:
\be
\S_2=-\frac{tr}{2g^2}∫ dx\,d^8θ\,\sum_{n=2}^{∞}\frac{(-1)^n}{n}\int d^2y\,d^2y\'...\,d^2y^{(n-1)'}\frac{A_2\, A_2\'\,...\,A_2^{(n-1)'}}{(y^1-{y^1}\')\,...\,\left({y^1}^{(n-1)'}-y^1\right)}\label{CS2}
\ee
where $A_2$'s are evaluated at the boundary, effectively removing the $y^3-$dependence. Furthermore, equation \ref{still3} implies that $A_{1,2}$ do not depend on $y^3$ on-shell. This is the same as imposing $F_{23}=F_{31}=0$ and $A_3=0$ everywhere. We can even substitute these `remaining' equations of motion above in action \ref{CS2}, which completely removes the $y^3-$dependence of $A_2$'s. We also note that though the CS action as we started with is not gauge invariant, the resulting $♫$ action on the boundary space is gauge invariant under a familiar gauge transformation: $δA_{2}=\D_{2}λ.$

Finally, to connect the above construction with the usual harmonic action, we use a specific dreibein (${e_a}^m$) parameterizing the Wick-rotated coset SU(2)/U(1) constructed in \cite{DPHfH}:
\begin{align}
g&=\bpm t & y\\ \frac{t-1}{y} & 1\epm\bpm\exp{\i\frac{ϕ}{2}} & 0\\0 & \exp{-\i\frac{ϕ}{2}}\epm\label{gH}
\\
\Rightarrow\quad\begin{split}
d_{0}&=-2\i ∂_{ϕ}\\
d_{+}&=\exp{\i ϕ}\left[∂_y+\frac{1}{y}\left(t-1\right)\(t\,∂_t+2\i ∂_{ϕ}\)\right]\\
d_{-}&=\exp{-\i ϕ}y\,∂_t
\end{split}\label{eam1}
\end{align} 
where $\bar{y}→t=\frac{1}{1+y\bar{y}}$ and the subgroup U(1) acts on the right. We can now rewrite the above action in terms of `flat' connections and recover the well-known $♫$ SYM action:
\be
\S_{♫}=-\frac{tr}{2g^2}\int dx\,d^8\theta\sum_{n=2}^{\infty}\frac{(-\i)^n}{n}\(\prod_{k=1}^{n}\int\frac{ dy_k\,dt_k}{y_k}\)\frac{A_{1-}A_{2-}\,...\,A_{n-}}{y_{12}\,y_{23}\,...\,y_{n1}}.
\label{SHH0}
\ee
where $y_{12}=\(y_1-y_2+\frac{ε}{\bar{y}_1-\bar{y}_2}\)$ and the volume element is explicitly written in terms of `modified' stereographic coordinates for the coset described above.

Furthermore,  we could also use a different coset construction for the internal space that has a different generator as a subgroup and is a `contraction' of the earlier coset:
\begin{align}
g&=\bpm 1 & y\\ 0 & 1\epm\bpm\exp{\frac{ϕ}{2}} & 0\\0 & \exp{-\frac{ϕ}{2}}\epm\bpm 1 & 0\\ \bar{y} & 1\epm\label{gHp}\\
\Rightarrow\quad\begin{split}
d_{0}&=2 ∂_{ϕ}+2\bar{y}∂_{\bar{y}}\\
d_{+}&=\exp{ϕ}∂_y-\bar{y}^2∂_{\bar{y}}-2\bar{y}∂_{ϕ}\\
d_{-}&=∂_{\bar{y}}\,.
\end{split}\label{eam2}
\end{align}
We have to exchange $y^2↔y^3$ to see that the dreibein does satisfy the conditions of \ref{ec} at the boundary $y^2=0$ now. This gives us a `different' harmonic hyperspace in which the SYM action reads almost the same as above (\ref{SHH0}) except that the connection $A_-$ gets replaced with $A_0$ and the internal space has a different volume element. This internal 2-manifold has a degenerate metric (just $dϕ^2$) but the volume element is properly defined from the 3-manifold's volume element as $\bar{y}→0$ and is simply: $\exp{-ϕ}dy\,dϕ$.

\vspace{20pt}
\sect{Reduction from Harmonic Hyperspace to Projective}
We basically have to reduce the Wick-rotated 2D $y-$space of $♫$ to 1D $y-$space of $\check{Π}$. This was done in \cite{DPHfH} by Wick rotating the sphere $\(\mathrm{introduction\,of\,} t=\frac{1}{1+y\bar{y}}\)$ and going to one of the two boundaries ($t=0\,\&\,1$) of the resulting hyperbolic space parameterized by a single coordinate $y$. The integration over $y$ was defined as the usual contour integration and the derivation of $\check{Π}$ hypermultiplets from $♫$ ones was shown by integrating out the $t-$dependence. Though, it was not clear that the integration contour itself was invariant under finite SU(2) transformations, which can shift the singularities across it.

In this section, we revise the arguments and show that the choice of contour is invariant and the integration can be consistently defined. For that purpose, we choose the Wick-rotated coset SU(1,1)/U(1) ($\sim$ SO(2,1)/SO(2) $\sim$ RP²) as defining the 2D internal space of $♫$ for the rest of this section.

\vspace{15pt}
\subsect{Internal Space}
In stereographic coordinates, the projective plane RP² has a circular boundary that is given by $y\bar{y}=1$. It can be shown that it is invariant under the symmetry group SU(1,1) as follows: Given that $\bpm a & b\\c & d \epm ∈$ SU(1,1) and the group `metric' is $\bpm 1 & 0\\ 0 & -1\epm$, the matrix entries of the group element get related: $c=\bar{b}$ \& $d=\bar{a}$. Then, if $y→\frac{a y+b}{c y+d}$, it is easy to see that $y\bar{y}=1$ is an invariant. Thus, the usual contour integration definition over this boundary can be used for the $y$-coordinate in $\check{Π}$, where the $\bar{y}-$coordinate takes a fixed value and is redundant:
\be
∮\frac{dy}{2π\i}\,\frac{1}{y^{n+1}}=δ_{n,0}\,.
\ee

\bfig[h]\bc
\includegraphics[height=6.0cm]{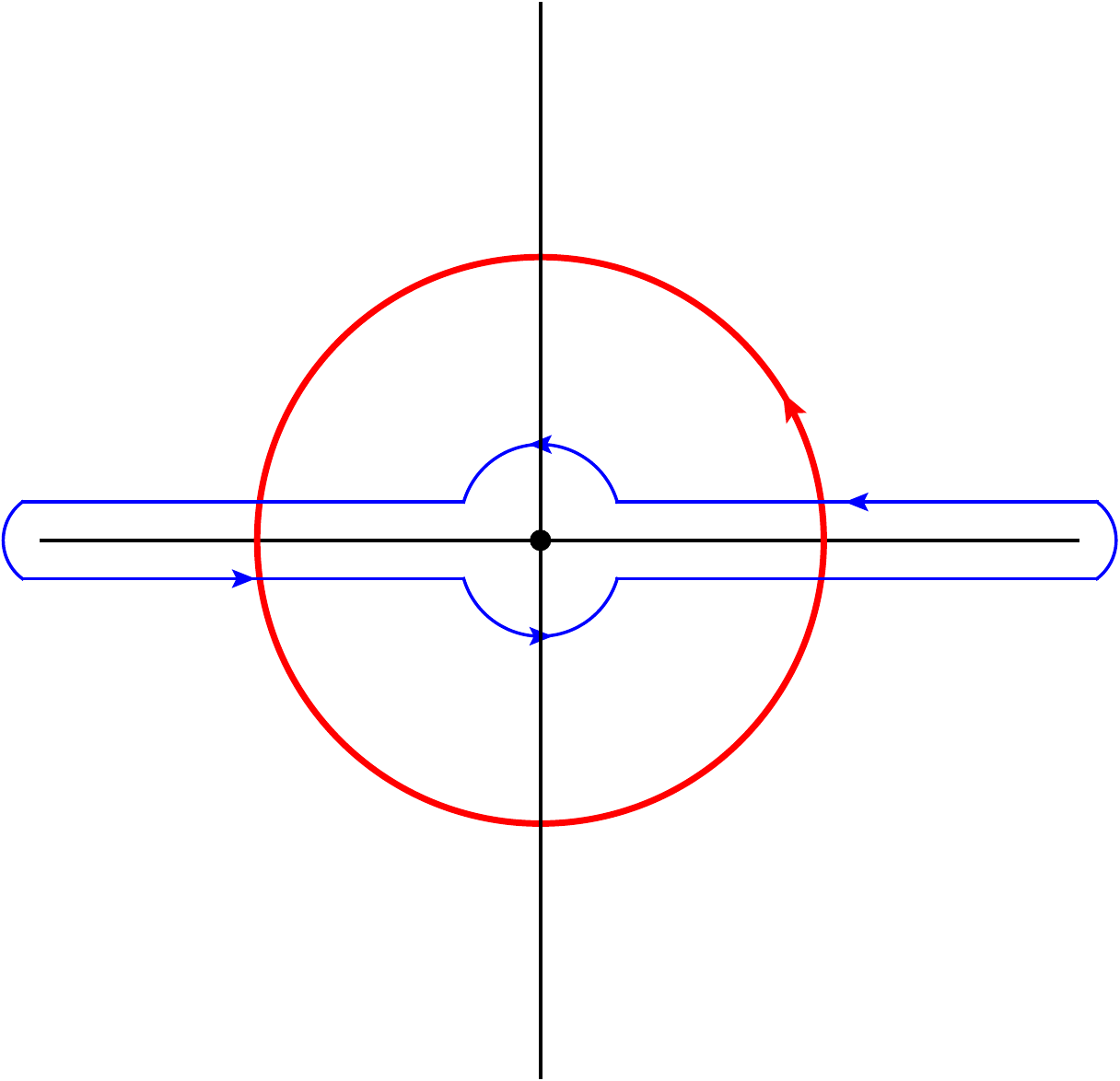}
\caption{Contours for $\check{Π}$ in $y-$plane.}
\label{Pcon}
\ec\efig

The same procedure still works if we Wick-rotate the isotropy group SO(2) to SO(1,1), which is optional at the level of harmonic hyperspace but required when reducing to the 1D internal space $(y)$ of $\check{Π}$ given by the coset SO(2,1)/SO(1,1)ISO(1). This can be achieved by `Wick-rotating' $\,\bar{y}→\frac{1}{\bar{y}}\,$ such that the boundary $y\bar{y}=1$ becomes $y=\bar{y}$, which is the `real' axis. This change basically corresponds to choosing an antisymmetric basis for the unitary `metric', i.e. $\bpm 0 & -\i\\ \i & 0\epm$ (which is usually chosen for SL(2,R) group) instead of the usual diagonal one as chosen above such that the modified group element now has purely real entries and reads (modulo the U(1)≡GL(1)-factor):
\begin{align}
g=&\,\frac{1}{\sqrt{\(1-y\bar{y}\)}}\bpm 1 & \bar{y}\\ y & 1\epm\nonumber\\
\xrightarrow{WR}&\,\frac{1}{\sqrt{\(1-\frac{y}{\bar{y}}\)}}\bpm 1 & \frac{1}{\bar{y}}\\ y & 1\epm\nonumber\\
\xrightarrow{CT}&\,\frac{1}{\sqrt{-2\i\bar{y}\(1-\frac{y}{\bar{y}}\)}}\bpm 1 & \frac{1}{\bar{y}}\\ y & 1\epm\bpm 1 & \i\\ \bar{y}  & -\i\bar{y}\epm\nonumber\\
⇒g=&\,\frac{1}{\sqrt{\frac{\i\(y-\bar{y}\)}{2}}}\bpm 1 & 0\\ \frac{y+\bar{y}}{2} & \frac{\i\(y-\bar{y}\)}{2}\epm.
\end{align}

The full transformation involves both the Wick-rotation and a coordinate transformation (CT). After this, the circular contour gets modified to a contour enclosing the `real' axis (see figure \ref{Pcon}) and effectively, the earlier definition of the contour integral can still be used by analytic continuation\footnote{If $y$ is to be treated as a complex coordinate, then this Wick-rotation is not required.}. This change now leads to transformation of the metric in stereographic coordinates to that in Poincaré coordinates and the corresponding volume elements read:
\be
\frac{dy\,d\bar{y}}{\(1-y\,\bar{y}\)^2}\,\xrightarrow{WR}\,\frac{dy\,d\bar{y}}{\(y-\bar{y}\)^2}\,.
\ee

\subsect{Action}
We now redefine $y≡y^1$ and $y^i=\{y^2,y^3\}≡\{\bar{y}\,(t),ϕ\}∈[0,1]$ to set the notation for projective hyperspace. We make a `special' Abelian gauge transformation for $A _y$:
\be
δA_y=∂_y\(∫^{y^i}_0 dy^{i'} A_{y^{i}}(y,y^{i'})\)≡∂_y λ
\label{AbA1}
\ee
where we assume $A_{y^i}|_{y^{i}=0}=0$. This relates the harmonic connection $A_y$ to the projective one as follows:
\begin{align}
♫:&\quad A_y=∂_y∫\frac{d^2y\'}{y\'-y}A_{y^i}\'\,;\quad \frac{1}{y\'-y}=P\(\frac{1}{y\'-y}\)+\i π δ(y-y\')θ(y^i-y^{i'})\label{y12def}\\
\check{Π}:&\quad A_y=∓\i\,∂_y∫\frac{dy\'}{y\'-y}V\'\,;\,\, V\'=±\i∫_{0}^{1}dy^{i'}\, A_{y^i}\'\quad\&\,\,\frac{1}{y\'-y}+\frac{1}{y-y\'}=\i πδ\(y\'-y\)
\label{Vdef}
 \end{align}
 Now, we can use this transformation to write down the action for Abelian SYM in projective space characterized by a 1D $y-$space:
\be
\S_{\check{Π}}^{(2)}=-\frac{tr}{4g^2}\int dx\,d^8\theta dy_1 dy_2\frac{V_{1}V_{2}}{y_{12}\,y_{21}}
\label{SPH2}
\ee
where $y_{12}$ is defined via equation \ref{Vdef} and the $ε-$prescription consistent with it reads $y_{12}=y_1-y_2+ε\(y_1+y_2\)$. This Abelian action is invariant under the following linear gauge transformation after identifying $λ|_{y^i=1}=Λ$ and $λ|_{y^i=0}=\bar{Λ}$:
\be
δ V=\i\(Λ-\bar{Λ}\).
\ee

The non-Abelian generalization of the definition of $V$ in eq. \ref{Vdef} is via path-ordered exponentiation that reads
\be
\exp{V}=\P\(\exp{\i∫_{0}^{1}dy^{i'}A_{y^i}\'}\).
\ee
This definition lifts the Abelian transformation of $V$ to the non-Abelian case as follows:
\be
\(\exp{V}\)\'=\exp{\i Λ}\exp{V}\exp{-\i\bar{Λ}}\quad⇒\quad δ\exp{V}=\i\(Λ\,\exp{V}-\exp{V}\bar{Λ}\).\label{nAgt}
\ee
The full non-Abelian SYM action that generalizes eq. \ref{SPH2} and is invariant under the non-Abelian gauge transformation of eq. \ref{nAgt} then reads (in close analogy to $♫$ action):
\be
\S_{\check{Π}}=-\frac{tr}{2g^2}\int dx\,d^8\theta\sum_{n=2}^{\infty}\frac{(-1)^n}{n}\(\prod_{k=1}^{n}\int dy_k\)\frac{\(\exp{V_{1}}-1\)...\(\exp{V_{n}}-1\)}{y_{12}\,y_{23}\,...\,y_{n1}}.
\label{SPH0}
\ee

\vspace{20pt}
\sect{Discussion \& Conclusion}
We have not been able to construct the projective covariant derivatives and field strengths, which would be the fundamental ingredients in the background field formalism for $\check{Π}$. However, we have an ansatz for the connection $A_y$ in terms of $V$ that comes very close to being the right one:
\be
A_y=\sum_{n=1}^{∞}(-1)^{n+1}\(\prod_{k=1}^{n}\int dy_k\)\frac{\exp{V}\(\exp{V_1}-1\)...\(\exp{V_{n}}-1\)}{\(y-y_{1}\)\,y_{12}\,...\,\(y_{n}-y\)}
\label{Ayexp}
\ee
because it produces the correct equation(s) of motion:
\be
d^4_ϑA_y=0 ⇒ d^2_ϑW=\bar{d}^2_ϑ\bar{W}=0.
\ee
However, $A_y$ in \ref{Ayexp} does not vary as a connection should, as can be checked with a straightforward calculation.
We expect that `regularizing' the divergent integrals by adding some projective terms should fix $A_y$ but we have not been able to find the correct pieces yet.

In conclusion, we have shown that a local CS action for N=2 SYM is equivalent to the usual action written in harmonic hyperspace $♫$. In fact, it seems that as long as consistent integration over the internal space of the harmonic formulation can be defined, the internal space need not be restricted to S² but can be spaces with boundaries like SO(2,1)/SO(2) or even degenerate spaces like its contraction SO(2,1)/ISO(1). We then showed that the 2D internal space(s) of the(se) harmonic hyperspace(s) when properly reduced to 1D reproduce the same projective hyperspace $\check{Π}$ as one would expect.

\section*{\bc\color{sect}{Acknowledgements}\ec}
This work is supported in part by National Science Foundation Grant No. PHY-0969739. DJ thanks Yu-tin Huang and Martin Ro\v{c}ek: Y-tH for pointing out reference \cite{CCDS} \& helpful discussions at the earlier stages of this work and MR for insightful discussions on $\check{Π}$ formalism.

\references{
\bibitem{GIKOS2-1}
A. Galperin, E. Ivanov, S. Kalitzin, V. Ogievetsky and E. Sokatchev, \cqg{1}{1984}{469};\\
A. Galperin, E. Ivanov, V. Ogievetsky and E. Sokatchev, {\it JETP\ Lett.\ } {\bf 40} (1984) 912 [{\it Pisma\ Zh.\ Eksp.\ Teor.\ Fiz.\ } {\bf 40} (1984) 155];\\
E. Ivanov, A. Galperin, V. Ogievetsky and E. Sokatchev, \cqg{2}{1985}{601}; \cqg{2}{1985}{617};\\
A.S. Galperin, E.A. Ivanov, V.I. Ogievetsky, and E.S. Sokatchev, {\it Harmonic superspace} (Cambridge Univ. Press, 2001).
\bibitem{Zup}
B. M. Zupnik, {\it Theor.\ Math.\ Phys.\ } {\bf 69} (1986) 1101 [{\it Teor.\ Mat.\ Fiz.\ } {\bf 69} (1986) 207].
\bibitem{DPHfH} 
D. Jain and W. Siegel, \pr{D}{80}{2009}{045024} [\arXivid{0903.3588} {\color{cyan}\small [hep-th]}].
\bibitem{ULMR}
U. Lindstr\"{o}m and M. Ro\v{c}ek, \cmp{115}{1988}{21}; \cmp{128}{1990}{191}.
\bibitem{WS10}
W. Siegel, 2010, \arXivid{1005.2317} {\color{cyan}\small [hep-th]};\\
D. Jain and W. Siegel, \pr{D}{83}{2011}{105024} [\arXivid{1012.3758} {\color{cyan}\small [hep-th]}].
\bibitem{GIKOS3-1}
A. Galperin, E. Ivanov, S. Kalitzin, V. Ogievetsky and E. Sokatchev, \cqg{2}{1985}{155}.
\bibitem{WEMSS}
E. Witten, \cmp{121}{1989}{351};\\ 
S. Elitzur, G. Moore, A. Schwimmer and N. Seiberg, \npb{326}{1989}{108}.
\bibitem{CCDS}
C-S. Chu and D. Smith, \jhep{01}{2010}{1} [\arXivid{0909.2333} {\color{cyan}\small [hep-th]}].
}
\end{document}